\documentclass{PoS}
\usepackage{amsmath,mathrsfs,bbm}

\title{Physics perspectives with heavy ions in the HL-LHC phase and beyond}

\ShortTitle{Physics perspectives with heavy ions}

\author{\speaker{Stefan Floerchinger}\\
        Institut f\"{u}r Theoretische Physik, Universit\"{a}t Heidelberg, 69120 Heidelberg, Germany\\
        E-mail: \email{floerchinger@thphys.uni-heidelberg.de}}

\abstract{The current state of research on high-energy heavy ion physics, including its motivations and purpose is reviewed from a theorist's perspective. Possible future directions are discussed, in particular the possibility of investigating the regime of small transverse momenta in more detail and an improved interplay between experiments and dedicated theory development.}

\FullConference{Sixth Annual Conference on Large Hadron Collider Physics (LHCP2018)\\
                4-9 June 2018\\
                Bologna, Italy}

\begin{document}

\paragraph{Introduction.} The discussion of the physics perspective for heavy ion collisions at the High Luminosity Large Hadron Collider (HL-LHC) is ongoing since some time, see for example refs.\ \cite{Jowett:2015dmf, Uras:2016njk, Antinori:2016zxe, Dainese, JFGO}. A CERN Yellow Report has been published on ``Physics at the FCC-hh, a 100 TeV pp collider'' including a chapter on ``Heavy ions at the Future Circular Collider'' \cite{Dainese:2016gch}, and another one on ``Heavy ions at the HL-LHC'' is currently in preparation. With the present contribution, I will not attempt to reflect, or summarize, the entire ongoing discussion, but rather present my own point of view (from a theorist's perspective). 

In general, heavy ion collisions, or the little bangs in the laboratory as they are sometimes called, pose a great challenge to modern physics. They allow to probe a quantum field theory at finite density and temperature, more precisely one of our fundamental gauge theories, namely QCD. The matter created by high energy nuclear collisions is strongly interacting and it undergoes interesting non-equilibrium dynamics. This makes the theoretical description challenging. At the same time, the whole field is strongly driven by experimental progress and as such a big motivation for further theory development. 

\paragraph{Fluid dynamics.} It has been learned in recent years, that much of the bulk dynamics of a heavy ion collision can be described in terms of {\it relativistic fluid dynamics} \cite{ALICE:2011ab, Chatrchyan:2012ta, ATLAS:2012at, Adare:2011tg, Adamczyk:2013gw}. Quite generally, it is believed that matter or quantum fields, when probed over sufficiently long distances and long time scales, and if interaction effects are strong enough, constitute a fluid. Fluid dynamics is a framework for an effective, macroscopic description of the dynamics and needs only a few {\it macroscopic fluid properties}. One of them is the thermodynamic equations of state, given for example in the grand-canonical ensemble by the pressure as a function of temperature and chemical potential, $p(T,\mu)$.\footnote{The chemical potential is here associated to net baryon number. In principle, one could introduce further chemical potentials associated to electric charge, strangeness and so on but this will be neglected here. For large collision energies such as at the LHC, even baryon chemical potential can be dropped for many purposes.} In addition, one needs transport properties such as shear viscosity, bulk viscosity, heat conductivity and additional second order terms such as relaxation times. In this context it is interesting to note that {\it ab initio} calculations of macroscopic fluid properties are difficult in practice because this demands control over the physics on many scales between the microscopic and the macroscopic domain. However, in principle they are fully fixed by the microscopic properties encoded in the Lagrangian of QCD.

The general idea underlying fluid dynamics is the belief that most (collective) degrees of freedom in a quantum field theory relax rather quickly towards their respective local equilibrium values. The slow modes are mainly those that are prevented from relaxing quickly by conservation laws. For a relativistic fluid, a central element in the formalism are therefore the covariant conservation laws for energy and momentum, $\nabla_\mu T^{\mu\nu}=0$, and for net baryon number, $\nabla_\mu N^\mu=0$. To make use of these, one decomposes the energy-momentum tensor and net-baryon number current as
\begin{equation}
T^{\mu\nu} = \epsilon  \, u^\mu u^\nu + ( p + \pi_\text{bulk}) \Delta^{\mu\nu} + \pi^{\mu\nu}, \quad\quad\quad N^\mu = n \, u^\mu + \nu^\mu.
\end{equation}
We use here the energy density $\epsilon$, the net-baryon number density $n$, the fluid velocity $u^\mu$ and the projector $\Delta^{\mu\nu}=g^{\mu\nu}+u^\mu u^\nu$. The pressure $p$ is related to $\epsilon$ and $n$ by the equilibrium relation encoded in the thermodynamic equation of state $p(T,\mu)$. The remaining terms are the bulk viscous pressure $\pi_\text{bulk}$, the shear stress $\pi^{\mu\nu}$ and the baryon diffusion current $\nu^\mu$. From the covariant conservation laws one obtains then evolution laws for energy density,
\begin{equation}
u^\mu \partial_\mu \epsilon  + (\epsilon + p + \pi_\text{bulk}) \nabla_\mu u^\mu + \pi^{\mu\nu} \nabla_\mu u_\nu = 0,
\end{equation}
for the fluid velocity,
\begin{equation}
(\epsilon+p+\pi_\text{bulk}) u^\mu \nabla_\mu u^\nu + \Delta^{\nu\mu}\partial_\mu (p+\pi_\text{bulk}) + \Delta^\nu_{\;\;\alpha}\nabla_\mu \pi^{\mu\alpha} = 0,
\end{equation}
and the baryon density,
\begin{equation}
u^\mu \partial_\mu n + n \nabla_\mu u^\mu + \nabla_\mu \nu^\mu = 0 .
\end{equation}
However, these equations are not closed because no relations have been provided for $\pi_\text{bulk}$, $\pi^{\mu\nu}$ and $\nu^\mu$ yet. This is actually an interesting point. In a non-relativistic fluid, described by the Navier-Stokes equation, one could simply express these ``dissipative stresses'' in terms of gradients of the fluid velocity, chemical potential and temperature with coefficients given by the bulk viscosity, shear viscosity and heat conductivity. For a relativistic fluid this is in fact not viable because it leads to equations of motion that are neither causal in the relativistic sense nor stable \cite{Israel:1979wp}! So at this point, relativistic fluid dynamics differs in an essential way from its non-relativistic counterpart. 

The way out was suggested by Israel and Stewart \cite{Israel:1979wp}. The dissipative stresses must be made dynamical fields. For example, the bulk viscous pressure can have an equation of the form,
\begin{equation}
\tau_\text{bulk} \, u^\mu \partial_\mu \pi_\text{bulk} + \pi_\text{bulk} + \zeta \; \nabla_\mu u^\mu + \ldots = 0.
\end{equation}
According to this equation, the bulk viscous pressure relaxes on a typical time scale of $\tau_\text{bulk}$ to the Navier-Stokes expression. The ellipses stand for higher order terms which are less important for the present discussion. Similar equations can be written down for the shear stress and baryon diffusion current. For a recent discussion of causality in the context of heavy ion collisions, see \cite{Floerchinger:2017cii}.

\paragraph{Thermodynamics.} One important ingredient for relativistic fluid dynamics is the thermodynamic information encoded in the equation of state. The pressure as a function of temperature $p(T)$ at vanishing baryon chemical potential is now understood rather well from lattice QCD calculations \cite{Borsanyi:2016ksw,Bazavov:2014pvz}. Also moments of conserved charges like e.\ g.\ for net baryon number,
\begin{equation}
\chi_2^B = \frac{\langle (N_\text{B} - \bar N_\text{B})^2 \rangle}{V T^3},
\label{eq:baryonnumbercumulant}
\end{equation}
are now understood rather well \cite{Bellwied:2015lba,Bazavov:2017dus}. Further progress can be expected during the next years, as a result of improved algorithms and also the steady progress in computing power.

\paragraph{Quantum fields and information.} Another interesting direction on the theoretical side is the connection between quantum field theory and (quantum) information theory. While such a relation is well understood in thermal equilibrium, a possible extension to out-of-equilibrium situations is currently an open question and topic of research. Interesting insights have been gained here from the AdS/CFT correspondence. In that framework, a quantum information theoretic characterization is accessible in terms of the entanglement entropy in an essentially geometric way \cite{Ryu:2006bv,Ryu:2006ef}. The AdS/CFT correspondence has already played a constructive role in understanding different phenomenological aspects of heavy ion collisions \cite{CasalderreySolana:2011us}. Although this theoretical framework cannot be applied to QCD directly (it needs a theory with $N_c\to \infty$), it has the big advantage that it can treat gauge theories in the strong interaction limit and out-of-equilibrium. 

Also for the direct understanding of heavy ion collisions in terms of non-equilibrium quantum field theory and relativistic fluid dynamics, information theoretic concepts could play an interesting role. Explorations in this direction are just starting; recently for example the role of quantum entanglement for hadron production from expanding QCD strings was investigated \cite{Berges:2017zws,Berges:2017hne}.

\paragraph{Fluctuating initial conditions.} A new discovery during the first years of heavy ion collisions at the LHC was about the importance of initial state fluctuations. It was understood since long that non-central collisions lead to an elliptic deformation of the initial energy density, to which the fluid dynamic response is a corresponding deformation of the azimuthal particle distribution called elliptic flow $v_2$. A quantitative definition is given by the following expansion,
\begin{equation}
\frac{dN}{d \phi} =\frac{N}{2\pi}\left[1 + 2 \sum_{m}\; v_m \;\cos\left(m\,(\phi - \psi_R)\right) \right].
\end{equation}
One could have expected a symmetry under rotations in the transverse plane $\phi\to \phi+\pi$, at least for symmetric nucleus-nucleus collisions. While this allows elliptic flow $v_2$ and other even flow coefficients $v_4$, $v_6$ and so on, it seems to exclude the odd coefficients $v_1$, $v_3$, $v_5$ etc. In studies of two-particle (and higher order) correlation functions at the LHC and at RHIC, such odd coefficients where found to be sizable, however \cite{ALICE:2011ab, ATLAS:2012at, Adare:2011tg}. It was then understood that they are present as a result of fluctuations in the initial transverse density distribution \cite{Alver:2010gr}. For example, in a Glauber-type model, the transverse positions of the nucleons that constitute a nucleus are fluctuating from event to event which leads to corresponding fluctuations of the energy density at the beginning of a fluid dynamic description.

\paragraph{Big bang - little bang analogy.} The fluctuating initial state opened a rather interesting avenue of research. In fact, the situation in a heavy ion collision resembles to some degree the one of cosmology. Also there, initial state fluctuations play an important role and understanding their evolution conceptually and quantitatively led to large steps forward in general understanding. Also the non-equilibrium, expanding character and the description by some form of fluid dynamics is shared between both systems. Obviously, there are also many differences. This starts from the typical scales - cosmological versus nuclear - but also includes the number of realizations - a single one versus very many - and the relevant forces. For heavy ion collisions everything is dominated by QCD, while for cosmology gravity and electromagnetism play a major role, together with the rather nebulous sector of dark matter and dark energy. 

Both systems share the complication that all information must be reconstructed from the final state and that initial conditions are not directly accessible to experimental investigation. Nevertheless, a lot could be learned (and still can be) from studying the various fluctuations and their evolution. A rather powerful theoretical tool is cosmological perturbation theory and a similar theoretical approach was also developed for heavy ion physics using a fluid dynamic description with a mode expansion \cite{Floerchinger:2013rya}.

There is another consideration that might be worth mentioning here. For high-energy nuclear collisions, a major task of the field is to understand the fluid properties, i.\ e.\ essentially the dynamics and correlation functions of the energy-momentum tensor, on the basis of known microscopic, or fundamental properties, in short $\mathscr{L}_\text{QCD}\to$ {\it fluid properties}. The situation is to some extend opposite for late time cosmology. Dark matter (and also dark energy) can only be accessed indirectly via gravitational interactions. According to Einstein's field equations, $G^{\mu\nu}=8\pi G_\text{N} T^{\mu\nu}$, one can in fact access only the energy-momentum tensor and its correlation functions. In the absence of any direct detection of dark matter, this is the only possibility to probe it. It goes without saying that one is ultimately interested in understanding the microscopic or fundamental properties of dark matter, so essentially: {\it fluid properties} $ \to \mathscr{L}_\text{dark matter}$. But this means that a good understanding of the relation between microscopic and macroscopic physics is needed here. Insights into this connection as they can be gained from heavy ion physics research are very valuable for this purpose.

\paragraph{Collective behavior in small systems.} Currently, an interesting debate in the field is concerned with collectivity in small systems. For heavy ion collisions, different versions of the harmonic flow coefficients $v_n$ (defined via multi-particle correlation functions and cumulants) have been taken as signs for collective effects and were interpreted in terms of a fluid dynamic expansion. Interestingly, the same observables look actually very similar for proton-nucleus and even proton-proton collisions of high enough final particle multiplicity! The interesting question is now: can the fluid approximation also work in something as elementary as a proton-proton collision? Or are there alternative explanations for these observations? Initial state effects such as gluon saturation at small Bjorken-$x$ have been proposed as a possible alternative; the debate is on-going.

\paragraph{Chemical freeze-out.} There are in fact many other very interesting research questions currently being discussed in the context of heavy ion physics. The physics of chemical freeze-out for example, is interesting because it seems here possible to test the dynamics of the chiral crossover transition from a gas of hadrons and resonances to the quark-gluon plasma. In the region of small baryon chemical potential, relevant to collider experiments at high energy, there are indications that the chemical freeze-out temperature agrees well with the chiral crossover temperature as calculated by lattice-QCD simulations \cite{Andronic:2017pug}. In the coming years, it will be interesting to test also moments and cumulants (or ratios thereof), such as the one in eq.\ \eqref{eq:baryonnumbercumulant}, and to test to which extend they can be described by thermal equilibrium calculations. 

\paragraph{Quarkonia.} Interesting research is also done on heavy quark flavor bound states such as the $\Upsilon$ and how they get modified in an heavy ion environment. For example, it has been demonstrated by CMS that all $\Upsilon$ states are suppressed, excited states even more so \cite{Khachatryan:2016xxp}. This avenue of research leads to better and better understanding of QCD bound states and also their interactions with a QCD medium.

\paragraph{Jet quenching.} Another interesting class of observables concerns jet quenching, i.\ e.\ the modifications of highly energetic partons when they traverse a medium such as a quark gluon plasma. Besides the dijet asymmetry in reconstructed transverse energies, $A_J = (E_{T1}-E_{T2})/(E_{T1}+E_{T2})$, with a separation in azimuthal angle $\Delta\Phi>\pi/2$, there are now many more detailed observables that allow to investigate how the jet structure gets modified in detail. In the future it should be possible (with better statistics) to investigate bottom quark jets and top quark jets and to clarify the interplay of microscopic partons and jets with the QCD fluid in more detail.

\paragraph{Light-by-light scattering.} An interesting test of fundamental QED physics was recently published by ATLAS \cite{Aaboud:2017bwk}. Ultra-peripheral ion collisions produce strong electromagnetic fields and in the equivalent photon approximation they can be seen as ``beams'' of quasi-real photons. ATLAS could now observe a few events of photon-photon or Halpern scattering $\gamma \gamma\to \gamma\gamma$, induced by non-linear QED effects. More detailed studies in this direction will become possible in the future. Another interesting direction for ultra-peripheral collisions are investigations of nuclear parton distribution functions.

\paragraph{Theory development.} A general observation for heavy ion physics is that many very interesting experimental results are available now or in reach for the close future. However, precise studies need an interplay of theory and experiment and it is the theory part that is often underdeveloped. To fully profit from the experimental capacities, more dedicated high quality theory development would be needed. For example, a ``standard model'' needs to be developed and maintained; this would allow to understand heavy ion collisions and QCD dynamics much better.

\paragraph{Higher energies.} An interesting question is how heavy ion collisions would look like at higher collisions energy, for example at the energies of the Future Circular Collider (FCC). This has been discussed in some detail in the CERN yellow report chapter on ``Heavy ions at the Future Circular Collider'' \cite{Dainese:2016gch}. Basically, the higher collision energy leads to a higher initial energy density and temperature at fixed Bjorken time $\tau$ after the collision. Also the total particle multiplicity $N_\text{ch}$ and the fluid volume at chemical and kinetic freeze-out will be somewhat higher. The larger volumes and particle multiplicities lead generically to better probes of collective physics, better statistic for hard probes and some new features such as the possibility of thermal charm quarks. For a more quantitative discussion see ref.\ \cite{Dainese:2016gch}. However, one should presumably not expect paradigmatic changes in the physics picture.

\paragraph{Low $p_T$ physics.} Another interesting front of future research might be the regime of low transverse momentum. In fact, advances in detector technology might allow to construct eventually a dedicated detector that allows to go down to $p_T\approx 10 \text{ MeV} \approx \frac{1}{20 \text{ fm}}$ or so. This is in fact an interesting scale of momenta because the corresponding de Broglie wave length corresponds to the typical extension of the fireball. It is also substantially below the scale of temperature at kinetic freeze-out $T_\text{fo} \approx 120 \text{ MeV}$. Probing this low $p_T$ regime is interesting for various reasons. One is that it can help to probe the macroscopic properties of the QCD fluid in more detail. Very soft pions, kaons, protons and in particular di-leptons carry additional information about the physics of thermalization and fluid dynamic transport that would be highly interesting to study. Also the dynamics of chiral symmetry restoration could be probed there in finer detail. As an example, a hypothetical condensate of pions as it may result from an unconventional production mechanism or from disoriented chiral condensates, could lead to an enhanced occupation number for pions in the low-$p_T$ range, more specifically at the scale of the inverse extension of such a condensate in the transverse plane. For di-leptons, it would be interesting to find the spectrum below the temperature scale, $p_T\ll T$, because one is then sensitive to the transport peak which is characterized by electric conductivity. A detector with low-$p_T$ coverage could lead to an excellent understanding of charmonia and bottomonia physics.

\paragraph{Conclusions.} In summary, high-energy nuclear collisions provide the highly interesting possibility to study a relativistic fluid governed by QCD. The experimental and phenomenological understanding of collective effects in heavy ion collision physics is steadily improving and there are currently also hints for collective physics in smaller collision systems such as proton-nucleus and proton-proton collisions. Future investigations will settle what precisely is going on there. 

There are highly interesting parallels between heavy ion collisions and different aspects of cosmology that motivate to study both of them in parallel. Theoretical models can partly be developed side by side. This concerns the more phenomenological level, for example to describe the evolution of fluid inhomogeneities in the linear and non-linear regimes, but also profound theoretical questions such as the connection between microscopic descriptions in terms of fundamental Lagrangians and more macroscopic effective fluid properties. This is a crucial prerequisite to understand the properties of dark matter on the basis of the fluid properties that are accessible to observation via gravitational interactions.

For the QCD fluid, or quark-gluon plasma, it should be possible to push the understanding to a deeper and more quantitative level, in a combined effort of dedicated future experimental and theoretical work.

\end{document}